\documentclass[aps,twocolumn,showpacs,floatfix]{revtex4}
\usepackage{graphicx}
\begin{document}

\title {Vortex nucleation in rotating Bose-Einstein condensates}
\author{G. M. Kavoulakis}
\affiliation{Mathematical Physics, Lund Institute of Technology,
                     P. O. Box 118, S-22100 Lund, Sweden}
\date{\today}

\begin{abstract}

We study the formation and stability of a single vortex state in a
weakly-interacting Bose-Einstein condensate that is confined in a
rotating harmonic potential. Our results are consistent with the
fact that any single off-center vortex is unstable. Furthermore, a vortex
state located at the center of the cloud first becomes locally
stable as the rotational frequency increases. Finally our study
implies the existence of hysteresis effects.

\end{abstract}
\pacs{PACS numbers: 03.75.Kk, 67.40.Vs} \maketitle

When rotated, a superfluid forms quantized vortex states. Numerous
studies have examined in the past vortices in the traditional
superfluid liquid Helium IV. Recently in some remarkable
experiments vortices have also been created and observed in vapors
of trapped ultracold atoms \cite{JILA, Madison, VortexLatticeBEC,
HaljanCornell}.

One of the basic questions in atomic systems is the formation and
stability of vortex states. Because of the confinement (which is
typically harmonic), there is a number of differences as compared
to homogeneous superfluids. For example, the energy spectrum is
discreet, the density is inhomogeneous and finally for harmonic
confinement the frequency of rotation of the gas is limited by the
trap frequency.

Many studies have examined the physics of vortices in trapped
gases \cite{DS, Dodd, Rokhsar, SF98, Svid98, Pu, IM, Garcia99,
Feder99, BR,PG,IM2}. The basic picture is that above a critical
frequency of rotation of the trap, one vortex state forms in the
gas, while as the rotational frequency increases further, more
vortices enter the cloud, eventually forming an array. One of the
most fundamental and important questions is thus the way that the
first vortex state forms.

In the present study we develop a method which allows us to
examine this problem in the limit of weak interactions, where the
typical atom-atom interaction energy is smaller than the
oscillator quantum of energy and one can restrict himself to the
subspace of states in the lowest Landau level. In this limit the
gas has a very peculiar property when the angular momentum per
atom ranges between zero and unity, as the interaction energy
scales linearly with the angular momentum \cite{BR, BP, KMP, JK,
KMR}, just like as in an ideal gas. This is an exact result within
the lowest Landau level subspace of states.

The linear behavior of the spectrum has important implications on
the rotational properties of the gas, as at a critical frequency
of rotation which is smaller than the trap frequency by a (small)
amount that is proportional to the ratio between the interaction
energy and the oscillator energy, the gas is predicted to undergo
a discontinuous phase transition from a non-rotating state to a
state with one vortex at the center of the trap \cite{BR}.

However, as we show below, one needs to be careful with the effect
of the truncation to the lowest Landau level wavefunctions. In the
present study, using second-order perturbation theory we calculate
the energy of the system to second order, where the energy no
longer increases linearly with the angular momentum. From the
derived dispersion relation we then examine the energy in the
rotating frame, finding that the gas exhibits a non trivial and
interesting behavior as function of the rotational frequency of
the trap $\Omega$.

Our study suggests that any single off-center vortex state is unstable.
Furthermore, a vortex state that is located at the center of the
cloud first becomes locally stable. We also predict that the gas
comes to rest at a smaller $\Omega$ than the one where rotation
sets in (see Fig.\,1), and the system exhibits hysteresis
\cite{PG}, which is a general characteristic of first-order phase
transitions. The graphs in Fig.\,1 show schematically the energy
of the gas in the rotating frame for certain physically-relevant
rotational frequencies which we calculate below.
\begin{figure}
\begin{center}
\includegraphics[width=7cm,height=3.9cm]{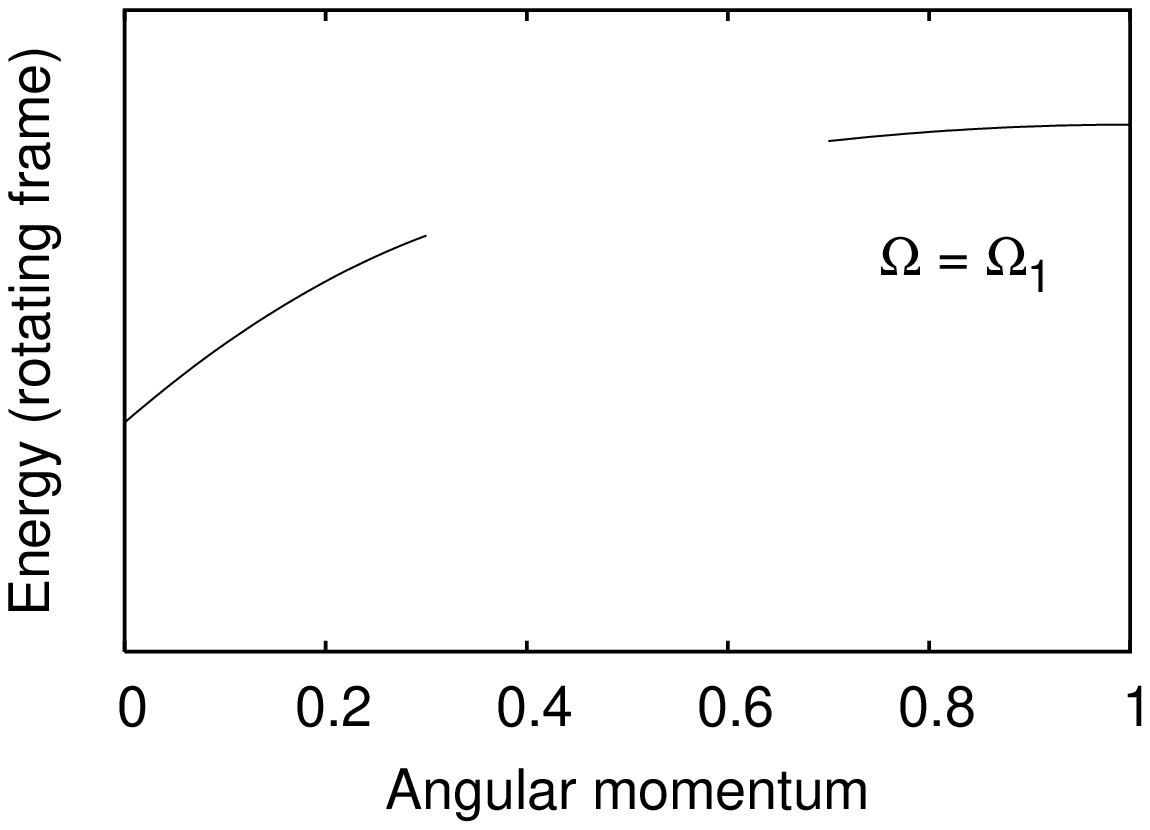}
\includegraphics[width=7cm,height=3.9cm]{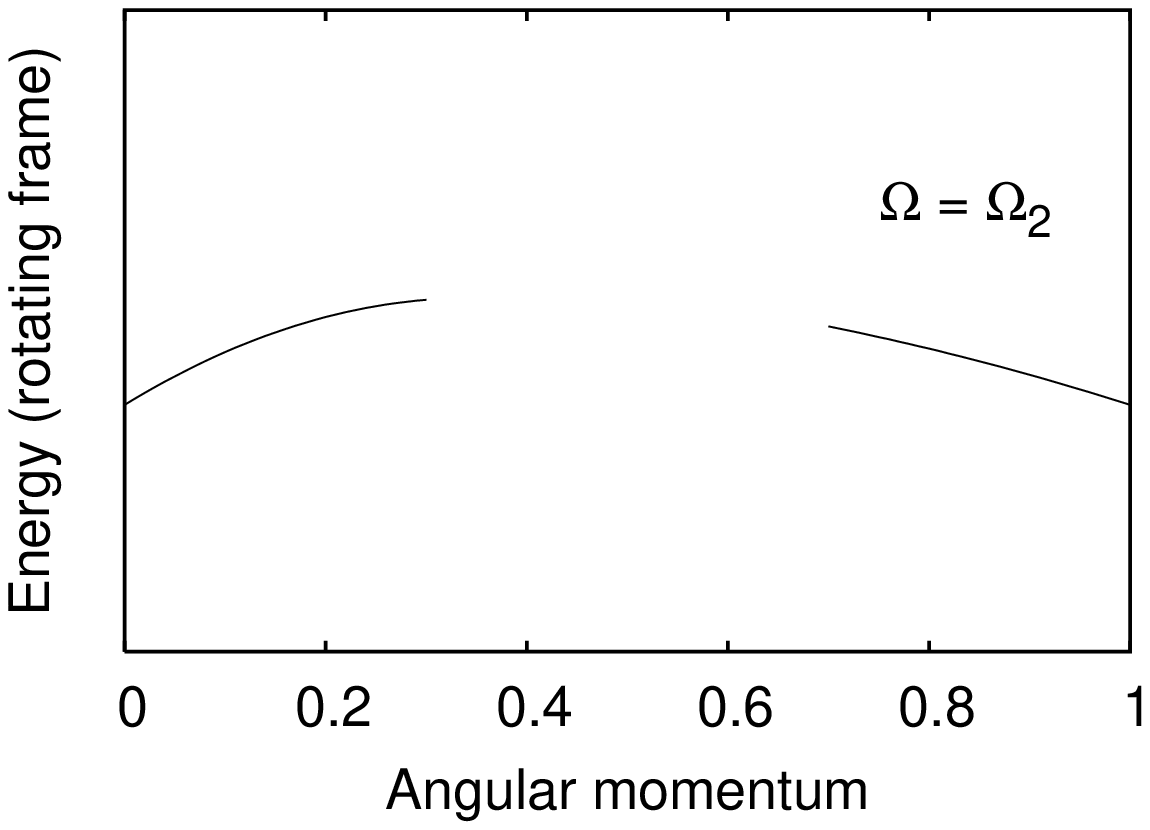}
\includegraphics[width=7cm,height=3.9cm]{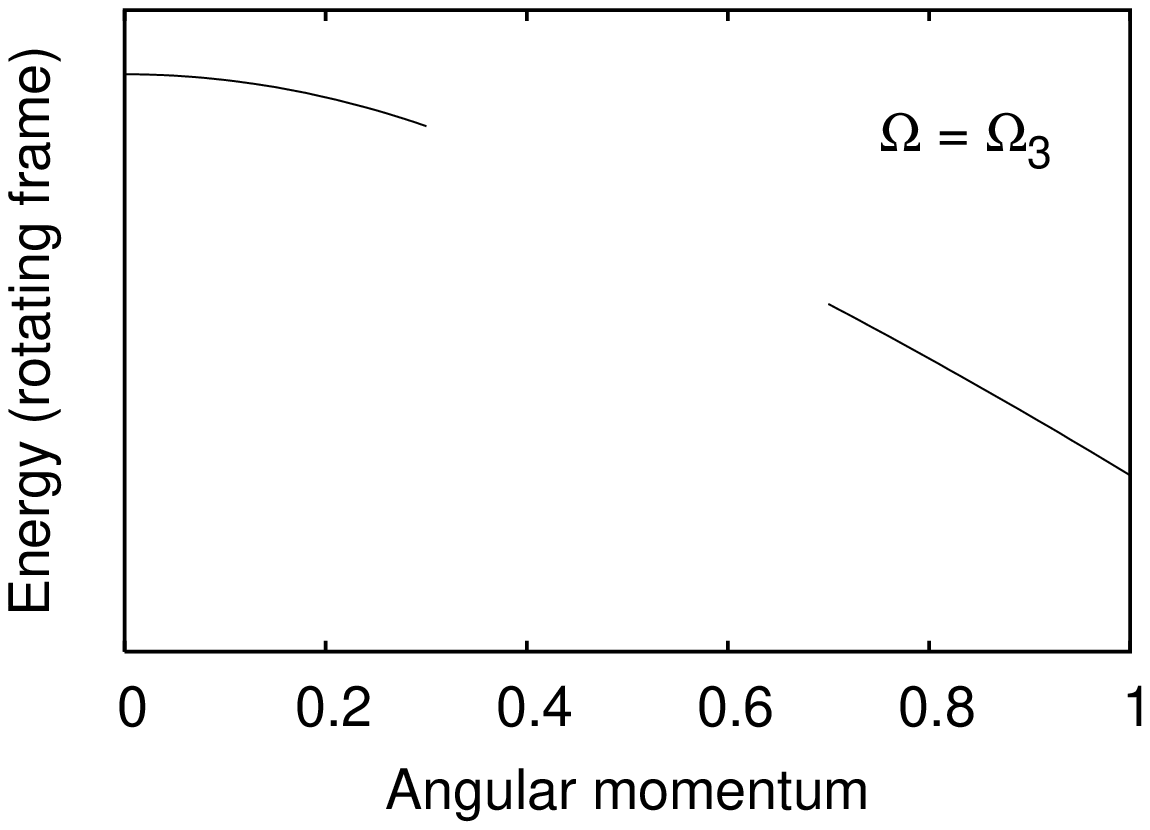}
\begin{caption}
{Schematic diagram which shows the energy of the gas in the
rotating frame as function of the angular momentum per atom $l$ for
different frequencies of rotation $\Omega$. On the top graph, for
$\Omega = \Omega_1$, the derivative of the energy at $l=1$ vanishes.
For $\Omega = \Omega_2$ the energy at $l=0$ equals that at $l=1$.
Finally, for $\Omega = \Omega_3$ the derivative of the energy at
$l=0$ vanishes. Our study shows that $\Omega_1 < \Omega_2 < \Omega_3$,
and also that the energy bents downward at both $l=0$ and $l=1$. All
these observations have crucial consequences on the formation and
stability of a single vortex state.}
\end{caption}
\end{center}
\label{FIG1}
\end{figure}

In our model we consider atoms interacting via a short-range
effective interaction,
\begin{equation}
    V_{\rm int} = U_{0} \sum_{i \neq j}
    \delta({\bf r}_{i} - {\bf r}_{j})/2.
\label{intlab}
\end{equation}
Here $U_0 = 4 \pi \hbar^2 a/M$ is the strength of the effective
two-body interaction, where $a$ is the scattering length for elastic
atom-atom collisions and $M$ is the atom mass. We also consider a
harmonic trapping potential of the form
\begin{equation}
V(\rho,z) = M \omega^2 (\rho^2 + \lambda z^2)/2,
\end{equation}
where $\rho$ and $z$ are cylindrical polar coordinates, $\omega$
is the oscillator frequency, and $\lambda$ is a dimensionless constant
($\hbar = M = \omega = 1$ from now on.)

We consider weak interactions and strong confinement along the $z$
axis, which is taken to be the axis of rotation. For weak
interactions the corresponding dimensionless quantity is $\gamma =
N a / d_z$ which is our expansion parameter as it is assumed to be
much smaller than unity. Here $N$ is the total number of atoms and
$d_z$ is the oscillator length along the axis of rotation. In this
limit one can work within the subspace of the nodeless ($n=0$)
eigenfunctions of the two dimensional harmonic oscillator,
\begin{equation}
  \psi_{n,m}(\rho, \phi) = \sqrt{ \frac{n!} {\pi (n+|m|)!}}
   \, \rho^{|m|} e^{i m \phi}
    L_n^m(\rho^2) e^{-{\rho}^2/2},
\label{basis}
\end{equation}
where $n$ is the number of radial excitations, $m$ is the quantum number
of the angular momentum, and $L_n^m$ are the associated Laguerre polynomials.
More specifically, the basis states are $\psi_{0,m}(\rho, \phi) =
\rho^{|m|} e^{i m \phi} e^{-{\rho}^2/2} / \sqrt{\pi |m|!}$.

The strong confinement along the axis of rotation, $\lambda \gg
1$, implies that the cloud is in its lowest state of motion along
this axis, and the problem thus becomes effectively two
dimensional, as the degrees of freedom along the $z$ axis are
frozen out. Therefore, the order parameter $\Psi({\bf r})$ can be
expanded in the product states $\Psi_{0,m}({\bf r}) = \psi_{0,m}({\rho,
\phi}) \, \phi_0(z)$, where $\phi_0(z)$
is the ground state of the one dimensional harmonic oscillator.

Under the above conditions, as shown initially by Bertsch and
Papenbrock \cite{BP} using numerical diagonalization, the
interaction energy of the gas in the lowest state is, for $2 \le L
\le N$,
\begin{equation}
   {\cal E}_{L,N} = \gamma (N - L/2 - 1) / \sqrt{2 \pi},
\end{equation}
where $L$ is the total angular momentum, and therefore it varies
linearly with $L$. References \cite{JK} have shown analytically
that this equation is {\it exact} to first order in $\gamma$.

Therefore, if $F = E - L \Omega$ is the total energy of the gas
in the rotating frame, where $E= L + {\cal E}_{L,N}$ is the total
energy in the rest frame then to first order in $\gamma$
(henceforth the energy is measured with respect to the energy of
the lowest state and terms of order $1/N$ are neglected), $F/N =
(1 - {\gamma}/2 \sqrt{2 \pi} - \Omega) \, l $, where $l = L/N$.
The above equation implies that the critical frequency for
rotation is
\begin{equation}
  \Omega_{c}^{(1)} = 1 -  \gamma / 2 \sqrt{2 \pi}.
\label{omc1}
\end{equation}
At this value of $\Omega$ Butts and Rokhsar \cite{BR} predict that
the gas undergoes a discontinuous transition from a non rotating
state to a state with a vortex located at the center of the trap.
In the present study we calculate the energy to next order for
$0 \le l \le 1$ examining how this picture is modified. On the
other hand, for $l > 1$ where more than one vortices are present,
calculation of the energy to lowest order in $\gamma$ suffices
to determine their stability.

We use now second-order perturbation theory to calculate the
interaction energy to next order, i.e., to $\gamma^2$. In a
similar method Ref.\,\cite{KMR} has calculated the low-lying
excitations of the system in the limit of low angular momentum. In
our study we use the results of Ref.\,\cite{KMP} which has studied
both regimes of low ($1 \ll L \ll N$) and high ($1 \ll N - L \ll
N$) angular momentum. Starting with $1 \ll L \ll N$, since we need
both the slope, as well as the curvature of the energy (which is
given by a term of order $l^{3/2}$ in this case), the order
parameter is, up to the desired order,
\begin{equation}
   \Psi = c_0 \Psi_{0,0} + c_2 \Psi_{0,2} + c_3 \Psi_{0,3},
\label{ext1}
\end{equation}
where $|c_0|^2 = 1 - l/2 + l^{3/2}/3$, $|c_2|^2 = l/2 - l^{3/2}$,
and $|c_3|^2 = 2 l^{3/2}/3$. The five diagrams which contribute to the
energy to second order in $\gamma$ and up to $l^{3/2}$ are shown in Fig.\,2.
\begin{figure}
\begin{center}
\includegraphics[width=5.5cm,height=2.6cm,angle=0]{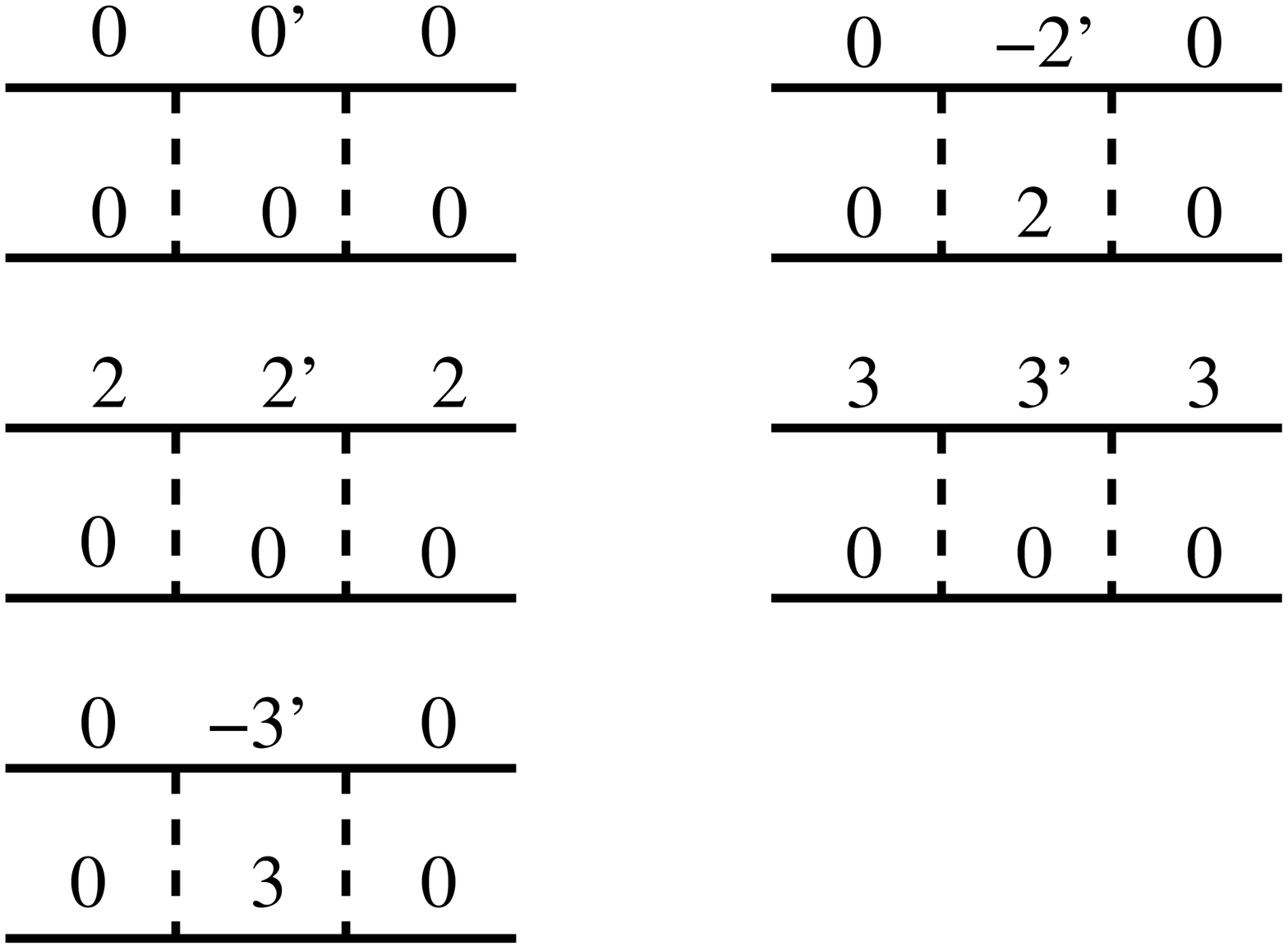}
\begin{caption}
{The five diagrams which contribute to the energy of the gas to
order $\gamma^2$ and up to $l^{3/2}$, for $1 \ll L \ll N$. The
unprimed (primed) numbers denote the angular momentum $m$ of the
states with zero (any nonzero) radial excitations, $n=0$ ($n \ge 1 $).
For the states with $m < 0$, $n$ can also be zero.
The dotted lines denote the interaction.}
\end{caption}
\end{center}
\label{FIG2}
\end{figure}
\begin{figure}
\begin{center}
\includegraphics[width=5.5cm,height=2.6cm,angle=0]{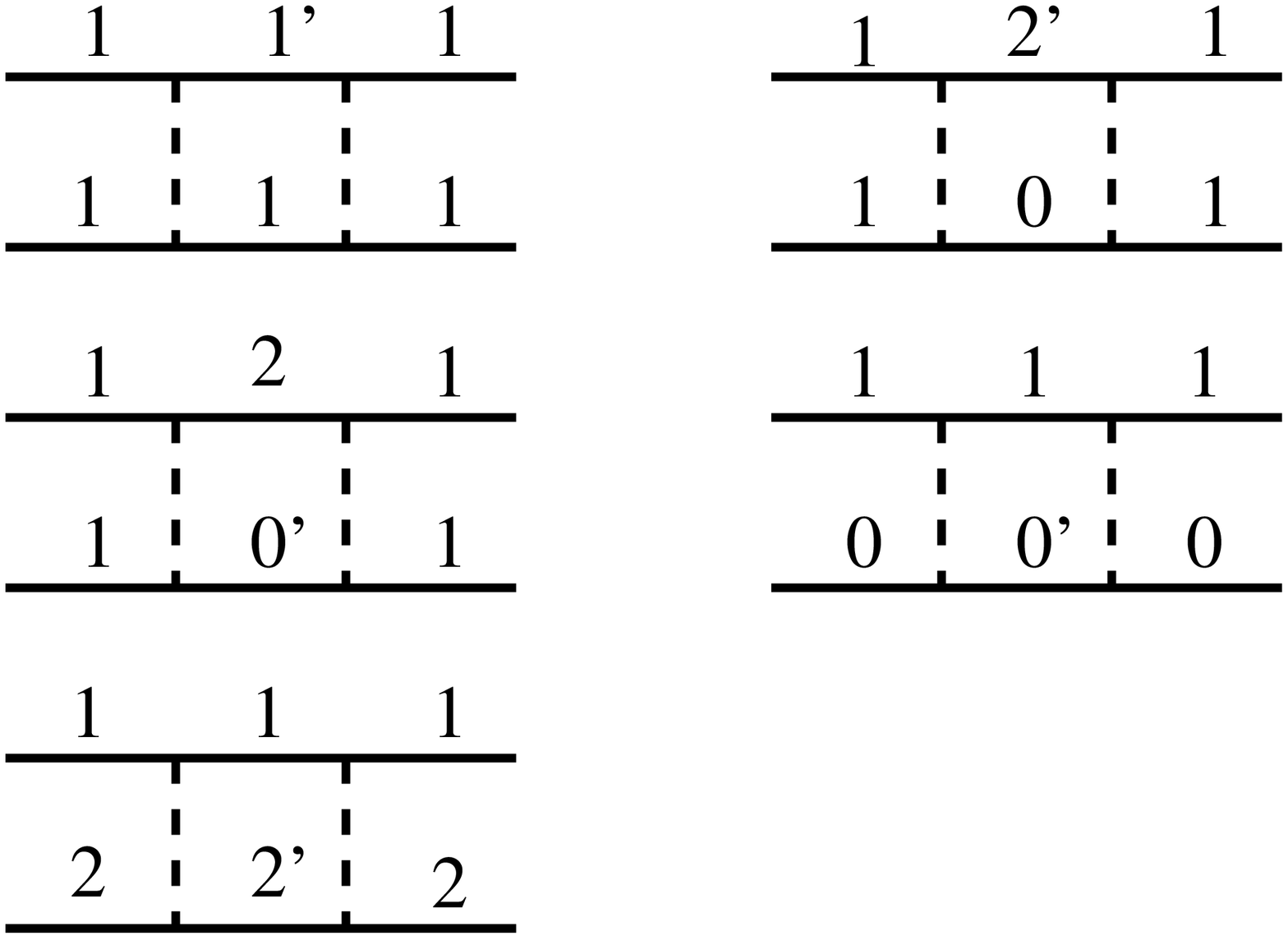}
\begin{caption}
{The five diagrams which contribute to the energy of the gas to
order $\gamma^2$ and up to $l^2$, for $1 \ll N - L \ll N $.}
\end{caption}
\end{center}
\label{FIG3}
\end{figure}
\begin{figure}
\begin{center}
\includegraphics[width=5.0cm,height=8.6cm,angle=-90]{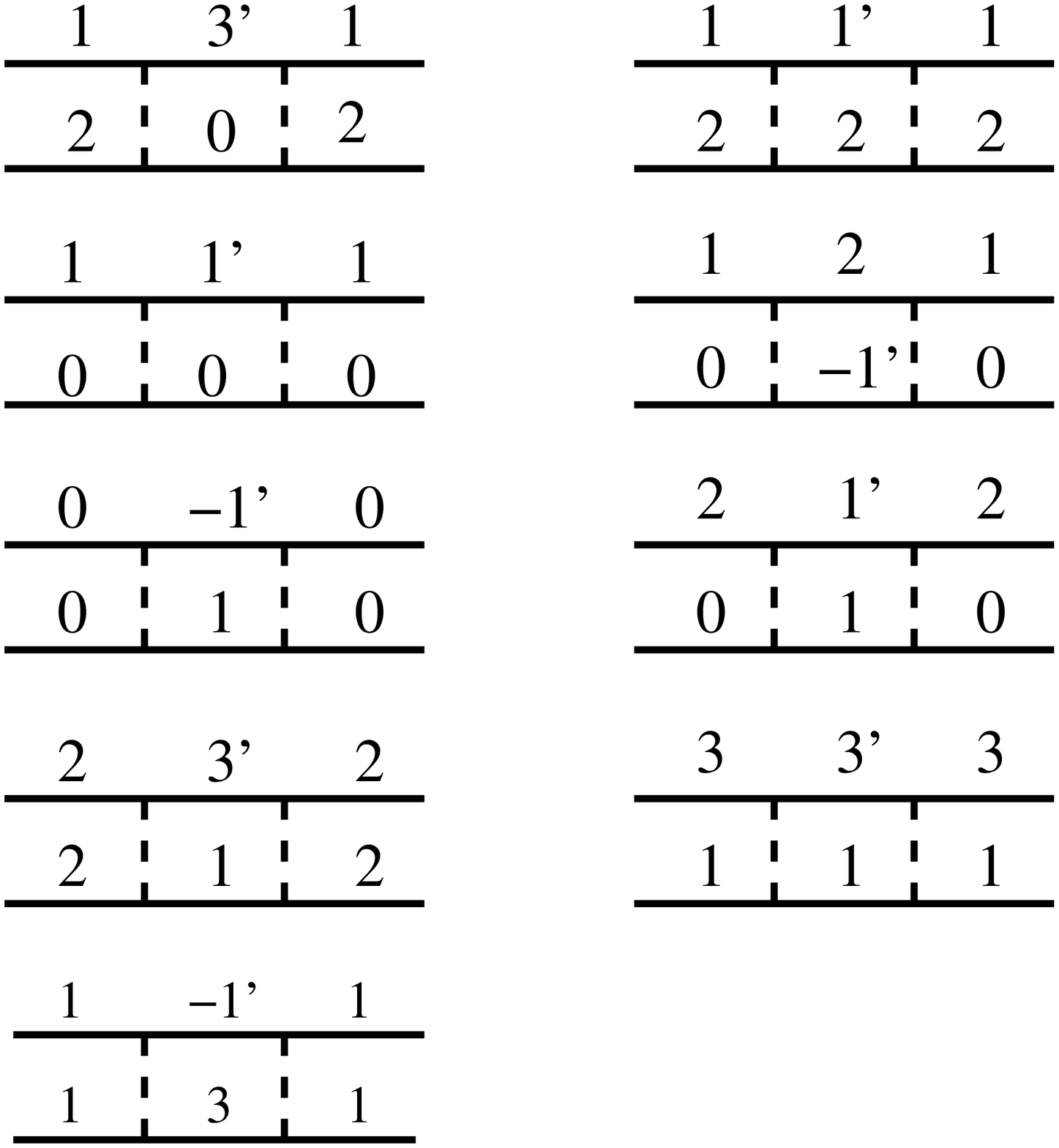}
\begin{caption}
{The nine diagrams which contribute to the energy of the gas to
order $\gamma^2$ and to $l^2$, for $1 \ll N - L \ll N $.}
\end{caption}
\end{center}
\label{FIG4}
\end{figure}
For example, the contribution of the top left diagram to the energy
per particle is equal to
\begin{eqnarray}
   \sum_{n=1}^{\infty}
   \frac {|\langle \Psi_{0,0}, \Psi_{0,0} | V_{\rm int} |
   \Psi_{0,0}, \Psi_{n,0} \rangle|^2}
   {N (\epsilon_{0,0} - \epsilon_{n,0})} =
    \sum_{n=1}^{\infty}  \frac {|I_{0,0}^{0,n}|^2 N_0^3 U_0^2}
   {N (\epsilon_{0,0} - \epsilon_{n,0})}
  \nonumber \\
  = - (1 - l/2 + l^{3/2}/3)^3 (N U_0)^2
  \sum_{n=1}^{\infty} \frac {|I_{0,0}^{0,n}|^2}
  {\epsilon_{n,0} - \epsilon_{0,0}},
\label{sum}
\end{eqnarray}
where $I_{0,0}^{0,n} = \int \Psi_{0,0}^* \Psi_{0,0}^* \Psi_{0,0}
\Psi_{n,0} \, d{\bf r}$ is the overlap integral between the
corresponding states. This always involves three states with zero
radial excitations and one state with $n$ radial excitation(s).
Also $\epsilon_{n,m} = 2 n + |m| + 1 + \lambda/2$ is the
eigenenergy of the states $\Psi_{n,m}({\bf r}) = \psi_{n,m}(\rho,
\phi) \phi_0(z)$ and $N_0 = N |c_0|^2$ is the occupancy of the
state $\Psi_{0,0}$. In Eq.\,(\ref{sum}) there is a factor of 1/2
from the interaction [Eq.\,(\ref{intlab})] that is cancelled by
a factor of 2 that comes from the symmetrization of the wavefunction.

The sum in Eq.\,(\ref{sum}) is over all the excited states with $m
= 0$ and $n \ge 1$ (in other cases where intermediate states with
negative values of $m$ are involved, $n \ge 0$.) However, the
series converges rapidly because of the overlap integrals which
decrease with increasing $n$. Considering the first fifteen
excited states, $n = 1, \dots, 15$ we find that
\begin{eqnarray}
 E/N \approx \left( 1 - \gamma/2 \sqrt{2 \pi} \right) \, l
 \phantom{XXXXXXXXXX} \nonumber \\
    - \left( 0.0916 - 0.0883 \, l + 0.0156 \, l^{3/2} \right) \gamma^2.
\label{fexc1}
\end{eqnarray}
As a final step we express $E/N$ as function of the expectation
value of the angular momentum per particle $\langle l \rangle
\equiv \langle {\tilde \Psi} | {\hat L} | {\tilde \Psi} \rangle /
\langle {\tilde \Psi} | {\tilde \Psi} \rangle$. Here ${\tilde
\Psi} = \sum_m c_m {\tilde \Psi_m}$, with ${\tilde \Psi_m} =
\Psi_{0,m} + \gamma \sum_{n \neq 0} d_{n,m} \Psi_{n,m}$ being the
perturbed basis states. The corrections of order $\gamma$ in the
basis states introduce a correction of order $\gamma^2$ in
$\langle l \rangle - l$.

To calculate $\langle l \rangle$ we start from the
Gross-Pitaevskii equation and make use of the orthogonality
between the states $\Psi_{n,m}$ finding $d_{n,m} = -2 \pi \int
|\Psi_{0,m}|^2 \, \Psi_{0,m} \, \Psi_{n,m}^* \, d{\bf r} /n$. This
formula implies that $\langle l \rangle = l [1 - 0.0405 \,
\gamma^2 - 0.0018 \, \gamma^2 \, l^{1/2}]$, or $l = \langle l
\rangle [1 + 0.0405 \, \gamma^2 + 0.0018 \, \gamma^2 \, \langle l
\rangle^{1/2}]$. Combining this result with Eq.\,(\ref{fexc1}) we
find
\begin{eqnarray}
 E/N \approx \left( 1 - \gamma/2 \sqrt{2 \pi} \right)
 \, \langle l \rangle \phantom{XXXXXXXXXX}
  \nonumber \\
      - \left( 0.0916 - 0.1288 \, \langle l \rangle + 0.0138 \,
      \langle l \rangle^{3/2} \right) \gamma^2.
\label{fexc1f}
\end{eqnarray}
From Eq.\,(\ref{fexc1f}) we conclude that the critical frequency
of rotation is, to order $\gamma^2$,
\begin{equation}
  \Omega_{c}^{(2)} \equiv \Omega_3 = 1 - \gamma/2 \sqrt{2 \pi} +
     0.1288 \, \gamma^2 + {\cal O} (\gamma^3).
\label{fnewf}
\end{equation}
Also $E/N$ (as well as $F/N$) has a downward curvature.

Following a similar method we perform the same analysis for $1 \ll
N - L \ll N$. There is no term of order ${\tilde l}^{3/2}$ in this
case and the two leading terms are of order ${\tilde l}$ and
${\tilde l}^2$, where ${\tilde l} = 1 - l$. The corresponding
order parameter is \cite{KMP}
\begin{equation}
   \Psi = c_0 \Psi_{0,0} + c_1 \Psi_{0,1} + c_2 \Psi_{0,2} + c_3 \Psi_{0,3},
\end{equation}
where $|c_0|^2 = 2 {\tilde l} - 3 {\tilde l}^2 /2$,
$|c_1|^2 = 1 - 3 {\tilde l} + 27 {\tilde l}^2/8$,
$|c_2|^2 = {\tilde l} - 9 {\tilde l}^2/4$, and
$|c_3|^2 = 3 {\tilde l}^2 /8$.

The five diagrams which contribute to the interaction energy
linearly and quadratically in ${\tilde l}$ are shown in Fig.\,3,
while the ones in Fig.\,4 contribute only quadratically. Again,
considering the first fifteen excited states we find
\begin{eqnarray}
   E/N \approx \left( 1 - \gamma / {2 \sqrt{2 \pi}} \right) (1 - {\tilde l})
\phantom{XXXXXXXXX} \nonumber \\
     - \left( 0.0111 + 0.1024 \, {\tilde l} + 0.7654 \, {\tilde l}^2 \right)
     \gamma^2.
\label{fexc12}
\end{eqnarray}
In this case $\langle {\tilde l} \rangle = {\tilde l} [1 + 0.0780
\, \gamma^2 - 0.0811 \, {\tilde l} \, \gamma^2]$, or ${\tilde l} =
\langle {\tilde l} \rangle [1 - 0.0780 \, \gamma^2 + 0.0811 \,
\langle {\tilde l} \rangle \, \gamma^2]$. The energy, expressed as
function of $\langle {\tilde l} \rangle$ is therefore
\begin{eqnarray}
   E/N \approx \left( 1 - \gamma / {2 \sqrt{2 \pi}} \right)
   \left( 1 - \langle {\tilde l} \rangle \right)
\phantom{XXXXXXX} \nonumber \\
           - \left(0.0111 + 0.0244 \, \langle {\tilde l} \rangle
        + 0.8465 \, \langle {\tilde l} \rangle^2 \right) \gamma^2.
\label{fexc12f}
\end{eqnarray}
The frequency at which the derivative vanishes is thus
\begin{equation}
  \Omega_1 = 1 - \gamma/2 \sqrt{2 \pi}
      + 0.0244 \, \gamma^2 + {\cal O} (\gamma^3).
\label{ffnewf}
\end{equation}
Again, $E/N$ (and $F/N$) has a downward curvature.

Finally, another relevant frequency of rotation is the one where
the energy in the rotating frame for $\langle l \rangle =0$ equals
that for $\langle l \rangle =1$, and this turns out to be
\begin{equation}
  \Omega_2 = 1 - \gamma/2 \sqrt{2 \pi}
      + 0.0805 \, \gamma^2 + {\cal O} (\gamma^3),
\label{ffnewf2}
\end{equation}
in agreement with Ref.\,\cite{LF}.

Having calculated the energy as function of the angular momentum
in the two limiting cases, we can extract valuable information
about the formation and stability of a single vortex state. Since
our Hamiltonian is rotationally invariant, it commutes with the
angular momentum and therefore the angular momentum is a good
quantum number. In that respect, any configuration is stable,
however the interesting question is the stability against weak
perturbations.

For example, in the presence of a small thermal component in the
gas which interacts with the condensate exchanging angular
momentum and energy with it, a vortex state is stable/metastable
as long as its energy in the rotating frame has an absolute/local
minimum \cite{TL}. In the limit of weak interactions that we
consider here, the energy is dominated by the oscillator energy
and the energy barriers are small. As a result, these systems
cannot support persistent currents [i.e., $F(l)$ does not have any
metastable minimum at any $l \neq 0$ when $\Omega = 0$]. Still,
when $\Omega \neq 0$, $F(l)$ develops in principle local minima
and in what follows, we consider a vortex state as a stable
configuration provided that $F(l)$ has a local/absolute minimum
(henceforth $l$ is to be identified as $\langle l \rangle$).

The angular momentum plotted on the horizontal axis of Fig.\,1 is
directly related to the position of the vortex, since for an
off-center vortex state $0 < l < 1$ \cite{PS}. Therefore, as $l$
increases in Fig.\,1 the vortex moves from an infinite distance
away from the trap ($l=0$) to its center ($l=1$).

Our results here are exact for weak interactions and for values of
the angular momentum close to zero and unity. Still, our expansion
strongly suggests that the schematic form of $F(l)$ shown in
Fig.\,1 extends over all the intermediate values of the angular
momentum, without any local minima in between.

As shown in Fig.\,1 for any frequency $\Omega$ between $\Omega_1$
and $\Omega_2$, $F(l)$ has a local minimum at $l=1$. In other
words a vortex that is located at the center of the trap first
becomes stable locally with increasing $\Omega$. This behavior is
qualitatively the same as in Ref.\,\cite{IM2} where the opposite
limit of strong interaction was considered.

According to our study, if one first cools down below the
condensation temperature and then rotates, the gas will undergo a
discontinuous transition from a non-rotating state to a state with
a vortex in the middle of the cloud for $\Omega = \Omega_3$. In
the reverse process as $\Omega$ decreases, the gas will make a
discontinuous transition from $l=1$ to $l=0$ at an $\Omega =
\Omega_1$, with $\Omega_1 < \Omega_3$.

In the other physically-relevant situation, if one first rotates
with some $\Omega = \Omega_0$ and then cools down below the
condensation temperature, the gas may actually reach the state
with a (stable) centered vortex for any value of $\Omega_0$
between $\Omega_1$ and $\Omega_3$.

It is crucial to mention that the slope of $F(l)$ increases
discontinuously as one crosses the point $l=1$ by an amount of
order $\gamma$, equal to $11 \gamma / 32 \sqrt{2 \pi}$ \cite{KMP}
(not shown in Fig.\,1). This fact guarantees that $F(l)$ increases
for $l \agt 1$ for all the values of $\Omega$ considered above.

To summarize, the spectrum of a weakly-interacting Bose-Einstein
condensate that is confined in a rotating harmonic trap implies
that a vortex state located at the center of the cloud becomes
first locally stable. On the other hand, any single off-center vortex is
unstable. Finally we predict hysteresis as the frequency of
rotation of the trap varies.

The author thanks Nikos Papanicolaou for suggesting this problem
to him and for useful discussions. He also thanks the Physics
department of the university of Crete for its hospitality.
This work was supported by the Swedish Research Council (VR),
and by the Swedish Foundation for Strategic Research (SSF).

\end{document}